\documentclass[epj]{svjour}

\usepackage{graphicx}
\usepackage{bm}
\usepackage{amsmath,amssymb}

\def\gs{\gtrsim}
\def\ls{\lesssim}
\def\be{\begin{equation}}
\def\en{\end{equation}}                  
\def\p{\partial} 
\newcommand{\bi}[1]{\mbox{\boldmath$#1$}}\def\bea{\begin{eqnarray}}
\def\ena{\end{eqnarray}}

\def\ge{> \kern -12pt \lower 5pt \hbox{$\displaystyle =$}}
\def\le{< \kern -12pt \lower 5pt \hbox{$\displaystyle =$}}
\def\gs{> \kern -12pt \lower 5pt \hbox{$\displaystyle{\sim}$}}
\def\ls{< \kern -12pt \lower 5pt \hbox{$\displaystyle{\sim}$}}

\def\ve{\varepsilon} 

\begin{document}
\title{Defect structures in  nematic liquid crystals 
around  charged particles}
\author{Keisuke Tojo\inst{1} \and Akira Furukawa\inst{2} 
\and Takeaki Araki\inst{1} \and Akira Onuki\inst{1} 
}                     
\offprints{}          
\institute{
Department of Physics, Kyoto University, Sakyo-ku, Kyoto 606-8502, Japan 
\and 
Institute of Industrial Science, University of Tokyo, Meguro-ku, 
Tokyo 153-8505, Japan
}

\date{Received: \today}

\abstract{
We  numerically 
study the orientation 
deformations in nematic liquid crystals around charged particles. 
We set up  a Ginzburg-Landau theory 
with inhomogeneous electric field. 
If the dielectric anisotropy 
$\varepsilon_1$ is positive, Saturn 
ring defects are formed around the particles.  
For $\varepsilon_1<0$,  novel ``ansa" defects appear, 
which are disclination lines 
with their ends on the particle surface. 
We find  unique defect structures  around two charged particles. 
To lower the free energy, oppositely charged particle pairs 
tend to be aligned in the parallel direction for $\varepsilon_1>0$ 
and in  the perpendicular plane for $\varepsilon_1<0$ 
with respect to the background director . 
For identically charged pairs the preferred directions 
for $\varepsilon_1>0$ and $\varepsilon_1<0$ are exchanged. 
We also examie competition between  the charge-induced 
anchoring and the short-range anchoring.  
If the short-range anchoring is sufficiently strong, 
it can be effective in the vicinity of the surface, 
while the director orientation 
is governed by the long-range electrostatic 
interaction far from the surface. 
\PACS{
{61.30.Dk}
{Continuum models and theories of liquid crystal structure}   \and
{61.30.Jf}
{Defects in liquid crystals}   \and
{77.84.Nh}{Liquids, emulsions, and suspensions; liquid crystals}   \and
{61.30.Gd}
{Orientational order of liquid crystals; 
electric and magnetic field effects on order } 
}
}
\maketitle

\section{Introduction}
A variety of  mesoscopic structures 
have been found  in  liquid crystals  
around inclusions such as colloids 
and water droplets 
 \cite{Review,Pou97,Zap99}. 
In nematics, inclusions  distort the orientation 
order over long distances, 
 inducing topological defects 
\cite{Lav00,Ter95,Lub98,P1,Allen,Yama,Hung,Araki1,Araki2,C}. 
We mention 
the formation of structures or phases, such as string-like 
aggregates \cite{Pou97,Araki2,Ska08,Rav07}, 
soft solids supported by a jammed cellular network of particles 
\cite{Mee00}, 
and  a transparent phase including microemulsions 
\cite{Yam01,Bel03}. 
The  origin of the long-range 
distortions  has been ascribed to the 
anchoring of the  liquid crystal molecules on the 
inclusion surface 
\cite{Lav00,Ter95,Lub98,P1,Allen,Yama,Hung,Araki1,Araki2,C,Fuk04,Sta04}. 
It arises from the  short-range molecular 
interactions between the  
liquid crystal molecules and the surface molecules. 
In  the Ginzburg-Landau-de Gennes  theory,  we  have 
a surface free energy depending 
on the orientation of liquid crystal molecules 
on the surface.

In this paper, 
 we are interested in  another anchoring mechanism. That is,  
electrically charged inclusions  align 
the liquid crystal molecules in their vicinity to lower 
the electrostatic energy \cite{NATO,Onu04,Foret06}, 
which can be relevant for ions and  charged 
particles. In fact,  de Gennes \cite{PG,Gen93}  
  attributed the origin of 
the small size of the ion mobility 
in nematics to a  long-range deformation  
of the orientation order around ions. 
 However, the effect of charges in liquid crystals remains 
complicated and has rarely been studied, 
despite its obvious fundamental and technological 
importance.  It is of great interest how the electric-field  anchoring 
mechanism works and 
how it is different from the usual short-range 
anchoring mechanism.

The  electric field and the 
liquid crystal  orientation are coupled because 
 the dielectric tensor $\ve_{ij}$ 
depends on the  local 
orientation tensor $Q_{ij}$ (see equation (14)). 
The alignment along a homogeneous electric 
field is well-known \cite{Gen93}, 
but  the alignment in an inhomogeneous electric 
field has not yet been well studied. 
When the dielectric tensor is inhomegeneous, 
it is a difficult task  to 
solve  the Poisson equation and seek  the 
electric potential $\Phi$. 
We here perform numerical simulations placing charged particles 
in liquid crystals  in a three-dimensional cell. 
We use the Ginzburg-Landau-de Gennes scheme in terms of  
the orientation tensor 
$Q_{ij}$ \cite{Foret06,Gen93,Scho,Hess}. 
A similar approach has recently been used to calculate 
the polarization and composition deformations 
around charged particles in electrolytes \cite{OnuKit04}. 
It is worth noting that hydration of water molecules around 
ions is analogous to the orientation anchoring 
of liquid crystal molecules around charged particles, 
as pointed out by de Gennes \cite{PG,Gen93}.

In Section 2, 
we will present a Ginzburg-Landau-de Gennes 
theory for liquid crystals containing charged particles. 
In particular, we will give  two general forms 
 of the electrostatic free energy 
for the fixed-charge and fixed-potential cases 
(which can be used for any dielectric fluids containg charges). 
In Section 3, we will 
explain the numerical method adopted in this work. 
in Section 4, we  will present numerical results 
of equilibrium configurations of the orientation order 
around charged particles. We will also examine competition 
of the short-range and electric-field  anchoring mechanisms. 
In Section 5, a summary and critical 
remarks will be given.

\section{Theoretical Background}

We consider a liquid crystal system in a cubic box 
and place one or two charged spherical particles 
with radius $R$ inside the box.   
The particle positions are written as 
 ${\bi R}_n$ ($n=1,2$).  
The liquid crystal order is described in terms of 
the symmetric  orientational tensor  
$Q_{ij}({\bi r})$ with the traceless condition 
$Q_{ii}=0$ \cite{Gen93}. 
We place one or two charged particles 
with radius $R$ considerably longer than the 
radius of the solvent molecules. 
In this work the Boltzmann  constant is set equal to unity 
and the temperature 
$T$ represents the thermal energy of 
a liquid crystal molecule.

\subsection{Model}

We are interested in the 
equilibrium liquid crystal orientation 
around the particles, which minimizes the sum of the 
 Landau-de Gennes free energy, the 
 short-range anchoring energy, and the electrostatic 
energy. Thus the total free energy of the liquid crystal containing 
charged particles consists of 
four parts as \cite{NATO,Onu04,Foret06} 
\be 
\mathcal{F}=\mathcal{F}_0+\mathcal{F}_g+\mathcal{F}_a+\mathcal{F}_e .
\en 
The first term  is of the Landau-de Gennes form, 
\bea
\mathcal{F}_0
=\int'  d\mbox{\boldmath $r$}\left[ 
\frac{A}{2} J_2-\frac{B}{3}J_3
+\frac{C}{4}J_2^2\right], 
\label{eq:F0}
\ena
where we introduce  
\be 
J_2=Q_{ij}^2, \quad 
J_3=Q_{ij}Q_{jk}Q_{ki}.
\en 
Hereafter repeated indices are implicitly summed over. 
The coefficient $A$ is dependent on the temperature $T$, 
while the coefficients $B$ and $C$ are positive constants 
assumed to be independent of $T$. The second term 
is  the gradient free energy in the 
 one-constant approximation, 
\bea
\mathcal{F}_g=\frac{L}{2}\int' d\mbox{\boldmath $r$}(\nabla_k Q_{ij})^2, 
\label{eq:Fg}
\ena
where $\nabla_k=\p/\p x_k$ ($x_k=x,y,z$) are the space derivatives 
and  $L$ is a positive constant.   
The space integrals  
$\int' d\mbox{\boldmath $r$}$ in equations (2) 
and (4) are  to be performed 
 only outside the particles 
$|{\bi r}-{\bi R}_n|>R$. It is  convenient to  define the length, 
\be 
d= T/L, 
\label{eq:d}
\en   
which is the typical molecular 
size of  liquid crystal. 
The  term $\mathcal{F}_a$ represents 
the short-range anchoring free energy. It  is  
expressed as the integral on  the particle surfaces, 
\be 
\mathcal{F}_a =   -w \int da {\nu}_i{\nu}_j Q_{ij},
\en   
where $da$ is the surface element,  
$\bi \nu$ is   the outward normal unit vector to the surface, and  
$w$ represents the  strength of the anchoring. 
For  the uniaxial 
form $Q_{ij}=S(n_in_j-\delta_{ij}/3)$, we have 
$\mathcal{F}_a= 
wS \int da  [1/3- ({\bi \nu}\cdot{\bi n})^2]$. 
Thus, for neutral particles, positive and negative values of $w$ 
lead to   homeotropic and planar anchoring, 
respectively.

\begin{figure}
\begin{center}
\includegraphics[width=0.45\textwidth,bb=0 0 258 117]{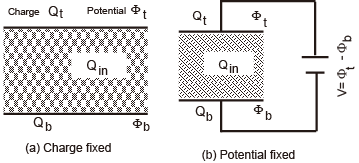}
\end{center}
\caption{A capacitor and an inhomogeneous  fluid  
containing a net charge $Q_{\rm in}$ 
in the fixed-charge case (a) 
and in the fixed-potential case (b). The charge and 
potential of the lower  plate are $Q_{\rm b}$ and $\Phi_{\rm b}$, 
 while those of the upper   plate are $Q_{\rm t}$ and $\Phi_{\rm t}$. 
}
\label{fig11}
\end{figure}

We  explain   the 
electrostatic part $\mathcal{F}_e$, which depends 
on the experimental method. 
As a generalization of the theory 
by one of the present authors \cite{NATO}, 
  we allow that the fluid region 
can contain a net charge $Q_{\rm in}=\int d{\bi r}\rho$, 
where $\rho=\rho({\bi r})$ is the charge density inside the fluid. 
As in Figure 1, we insert the fluid between 
  parallel metallic plates in the region $0<z<H$. 
The surface charge and the potential 
of the lower  plate at $z=0$ are $Q_{\rm b}$  and $\Phi_{\rm b}$, 
while those of the upper 
 plate at $z=H$ are $Q_{\rm t}$  and $\Phi_{\rm t}$. 
 We require  the overall  charge 
neutrality condition,  
\be 
Q_{\rm in}+Q_{\rm b}+Q_{\rm t}=0, 
\en
since the eletric field in the metal plates should vanish.    
In terms of  $Q\equiv (Q_{\rm t}-Q_{\rm b})/2$, 
we may set 
\be  
Q_{\rm b}=-Q -Q_{\rm in}/2, \quad 
Q_{\rm t}=Q -Q_{\rm in}/2.
\en 
(i) In (a)  in Figure 1, $Q$ can be fixed and can 
be  a control parameter, where  the potential difference,
\be 
V= \Phi_{\rm t}-\Phi_{\rm b},
\en 
 depends on the fluid inhomogeneity induced by the chaged particles. 
Here  the electrostatic energy of the 
surface charges of the plates 
is fixed, the appropriate form of $\mathcal{F}_e$ is 
\bea 
\mathcal{F}_e
&=& \frac{1}{8\pi}\int d{\bi r}{\bi D}\cdot{\bi E}\nonumber\\
&=&\int d{\bi r}\frac{\rho}{2}\bigg
(\Phi-\frac{\Phi_{\rm t}+\Phi_{\rm b}}{2}\bigg)+
\frac{QV}{2},
\ena 
where $E_i=-\partial \Phi/\partial x_i$ is 
the  electric field and ${D}_i=\ve_{ij}E_j$ 
is the electric induction with 
$\ve_{ij}$ being the dielecric tensor. 
Here we superimpose  small variations  
$\delta Q_{\rm b}$, $\delta Q_{\rm t}$, $\delta \rho$,  
 and $\delta \ve_{ij}$ 
on $Q_{\rm b}$, $ Q_{\rm t}$, $\rho$, 
and $\ve_{ij}$, respectively. 
We use the relation  $\int d{\bi r}{\bi E}\cdot\delta{\bi D}/4\pi=
\Phi_{\rm b}\delta Q_{\rm b}+\Phi_{\rm t}\delta Q_{\rm t}
+ \int d{\bi r}\Phi \delta\rho$. 
We then  obtain the incremental change of $\mathcal{F}_e$ as 
\bea 
\delta\mathcal{F}_e&=&V\delta Q +  \int d{\bi r}{\delta \rho}
\bigg
(\Phi-\frac{\Phi_{\rm t}+\Phi_{\rm b}}{2}\bigg)\nonumber\\
&& -\frac{1}{8\pi}\int d{\bi r}{\delta\ve_{ij}} E_iE_j.
\ena 
(ii) On the other hand, in (b)  in Figure 1, 
the potential difference $V$ 
can be fixed and can be a control parameter with 
$Q$ being dependent on the fluid inhomogeneity. 
The appropriate form of $\mathcal{F}_e$ is 
\bea 
\mathcal{F}_e&=&\frac{1}{8\pi}
\int d{\bi r}{\bi D}\cdot{\bi E}- VQ
\nonumber\\
&=&
\int d{\bi r}\bigg[\rho \bigg(\Phi-\frac{\Phi_{\rm b}+\Phi_{\rm t}}{2}\bigg) 
-\frac{{\bi D}\cdot{\bi E}}{8\pi}\bigg],
\ena 
where the second line follows from the second line of 
equation (10). This is the Legendre transformation of 
the electrostatic free energy in the fixed-charge case. 
Here we use  the same notation $\mathcal{F}_e$ 
in the two cases. Then  
 the incremental change of  $\mathcal{F}_e$ reads 
\bea 
\delta\mathcal{F}_e&=&-Q\delta V +  \int d{\bi r}{\delta \rho}
\bigg
(\Phi-\frac{\Phi_{\rm t}+\Phi_{\rm b}}{2}\bigg)\nonumber\\
&& -\frac{1}{8\pi}\int d{\bi r}{\delta\ve_{ij}} E_iE_j.
\ena 
where  the first term on 
the right hand side is different from that 
in equation (11). 
It is worth noting that the second line of 
equation (12) yields 
the frequently used 
expression $\mathcal{F}_e=-
\int d{\bi r}{\bi D}\cdot{\bi E}/8\pi$ 
in the fixed-potential condition 
for dielectric fluids 
without charge ($\rho=0$) 
(see reference\cite{PG}, for example).

The potential $\Phi$ 
satisfies the Poisson equation, 
\bea
\nabla_i (\ve_{ij}\nabla_j \Phi)=-4\pi\rho. 
\label{eq:Poisson} 
\ena
We assume the linear form of the dielectric tensor,  
\bea
\ve_{ij}(\mbox{\boldmath $r$})
=\ve_0\delta_{ij}+\ve_1Q_{ij}(\mbox{\boldmath $r$}),  
\label{eq:epsilon}
\ena
in the  liquid crystal  region (the particle 
exterior) \footnote{In the nematic state 
we have $\ve_{\parallel}=\ve_0+2S\ve_1/3$ 
along the director $\bi n$ 
and  $\ve_{\parallel}=\ve_0-S\ve_1/3$ 
in the perpendicular directions \cite{PG}, 
where the amplitude 
$S$ is given in equation (20).}.  Defining 
 $\Phi$ in the whole  space, we  may solve equation 
(14) by setting 
$
\ve_{ij}(\mbox{\boldmath $r$})=\ve_p\delta_{ij}
$ 
in  the particle interior. Then the integrals in equations 
 (10) and (12) 
are  over the whole cell region. 
We then have 
$\delta { \mathcal F}_{e} /\delta Q_{ij} =
-{ \ve_1} E_iE_j/8\pi$ 
both at fixed $Q$ and at fixed $V$.

\subsection{Equilibrium conditions}

In our numerical work we will adopt  the geometry 
(b) in Figure 1 and set $V=0$.
 The charge density   $\rho$ is fixed. 
We define the tensor, 
$h_{ij}\equiv 
 {\delta \mathcal{F}}/{\delta Q_{ij}}+\lambda\delta_{ij}$, 
where $\lambda$ is chosen such that $h_{ij}$ becomes traceless.  
Some calculations give  
\bea 
h_{ij} 
&=&(A+CJ_2)Q_{ij} 
-B \left(Q_{ik}Q_{kj}-\frac{1}{3}J_2\delta_{ij}\right)\nonumber\\
&-& L \nabla^2Q_{ij}  -\frac{ \varepsilon_1}{8\pi}
\left(E_i E_j-\frac{1}{3}{E}^2\delta_{ij}\right).  
\ena 
In equilibrium,   minimization of   
${\mathcal F}$ yields 
\be 
h_{ij}=0,
\en 
in the particle exterior. 
The  boundary condition of $Q_{ij}$ 
on the particle surface is given by  
\be
L\mbox{\boldmath $\nu$} 
\cdot \nabla Q_{ij}+{w}(\nu_i\nu_j-\delta_{ij}/3)=0.  
\en
 Obviously,  
the defect structure is independent of the sign of 
the particle charge, 
since $Q_{ij}$ is coupled to the bilinear terms of $\bi{E}$ 
in equation (16).

For $B>0$  uniaxial 
states  are selected in the bulk region below 
the isotropic-nematic transition 
   $A<A_{\rm t}$ \cite{Gen93}, 
where   $Q_{ij}=S(n_in_j-\delta_{ij}/3)$ 
and 
\be 
A_{\rm t}= B^2/27C. 
\en
Substituting  the uniaxial form 
into  the first line of equation (16), 
we  obtain   
${2C}S^2-{BS}+3A=0$, which is solved to give    
\be 
S= B/4C +[(B/4C)^2-3A/2C]^{1/2}. 
\en    
Just below  the transition 
we have $S=S_{\rm t}\equiv B/3C$. 
However, it is known that the liquid crystal order 
is considerably biaxial inside defect 
cores \cite{Allen,Foret06,Scho}. 
See Figure 3 of 
Ref.\cite{Foret06} for the biaxiality 
of the  Saturn ring core (where 
the spatial mesh size 
is  finer than in this work).  
Note that $Q_{ij}$ can generally be expressed as  
\be 
Q_{ij}=S_1(n_in_j-\delta_{ij}/3)+S_2(m_im_j-\ell_i\ell_j),
\en 
 where $\bi n$, $\bi m$, and $\bi \ell$ constitute 
three orthogonal unit vectors. 
Inside defect cores, the amplitude $S_2$ of biaxial order 
is of the same order as  the amplitude 
$S_1(=S$ in this work) of uniaxial order. 
Outside the defect cores, $S_2$ nearly vanishes 
and the orientation order becomes  uniaxial.

In addition,  the polarization vector 
of the liquid crystal is given by 
$P_i= \chi_{ij}E_j$ in terms of 
the susceptibility tensor $\chi_{ij}$. From $\ve_{ij}E_j= 
E_i+4\pi P_i$, we have  
\be 
\chi_{ij}= (\ve_{ij}-\delta_{ij})/4\pi.
\en 
This tensor should  be positive-definite in equilibrium 
to ensure 
 the thermodynamic stability 
in the (paraelectric) nematic  phase \cite{Foret06}. 
For the special form (15) this requirement 
becomes   
\be 
\ve_0-1+ \ve_1 q_\alpha>0,
\en 
where 
$q_\alpha$ ($\alpha=1,2,3$) 
are the eigenvalues of $Q_{ij}$.

\subsection{Electric field effect near the surface}

Let us consider the 
electric field effect near a 
particle surface. For simplicity 
we assume $|\ve_1|\ls \ve_0$. Then  
the surface electric field  $E_s$ is estimated 
to be of order  $eZ/\ve_0R^2$, where   
$Ze$ is the particle charge  (with $e$ being the 
elementary charge). 
(i) Far above the transition 
$A\gg A_{\rm t}$ in the isotropic phase, we neglect the terms 
proportional to $B$, $C$, and $L$  in equation (17) 
to obtain  
$Q_{ij} \cong \ve_1 E^2(x_ix_j/r^2-\delta_{ij}/3)
/8\pi A$, 
 which grows as $A$ is decreased as a pretransitional effect.  
(ii) Just below the transition, a nonlinear  deformation occurs for  
$|\ve_1|E_s^2/8\pi \gs  A_{\rm t}S_{\rm t}= 
B^3/81 C^2$, which is easily realized for small $B$. 
(iii) In the nematic phase far below the transition, 
strong nonlinear deformations of $Q_{ij}$ 
 are induced  on the  surface for  
 $R<\ell$ with \cite{Foret06} 
\be 
\ell = |Z|(|\ve_1| \ell_B d /12\pi \ve_0 S)^{1/2},
\en 
where  $d$ is defined by equation (5) and  
\be 
\ell_B=e^2/\ve_0T
\en 
is the Bjerrum length. 
This criterion arises from the balance of 
the gradient term ($\sim LSR^{-2}$) 
and the electrostatic term ($\sim \ve_1E_s^2/8\pi\propto R^{-4}$) 
in $h_{ij}$ in equation (17). 
Furthermore, for sufficiently large 
$\ell/R$, a defect is  formed around the  
particle, where the distance from 
the surface is   of order $\ell-R$.

It is important to 
clarify the condition of defect formation 
in real systems. Let us assume 
 $\ve_0\sim 2$, $|\ve_1|\sim \ve_0$, 
$S \sim 1$, $d\sim 2$nm, and $\ell_B \sim 24$nm. 
Then $\ell \sim |Z|$nm. 
Thus, the relation $R<\ell$ 
 holds for microscopic ions, though 
our  coarse-grained model is 
inaccurate on the angstrom  scale.
See the remark (3) in the last section 
for a  comment on ions in liquid crystal. 
We may also consider a large particle   
with  a constant surface charge density 
\be 
\sigma = Z/4\pi R^2.
\en 
It may  be difficult to 
induce sufficient ionization 
on  colloidal surfaces    in  liquid crystal 
solvents.  One  method of realizing  
charged surfaces  will  
be to attach  ionic surfactant molecules on 
colloidal surfaces.   
For such a particle,  the  condition of defect 
formation becomes  $R\gg R_c$, where 
\be 
R_c= (3\ve_0S/4\pi |\ve_1|\ell_B d)^{1/2}\sigma^{-1}. 
\en  
Using the above parameter values, 
 we have $R_c \sim 0.1 \sigma^{-1}$nm (with $\sigma$ in units 
of nm$^{-2}$). For example, 
if $\sigma= 0.0624$nm$^{-2}$ 
or $e\sigma=1 \mu{\rm C/cm^2}$, we obtain 
$R_c=1.6{\rm nm}$. 
Here the electric field at the surface is 
$e\sigma/4\pi \ve_0 \sim  100{\rm V/\mu m}$, which 
 is strong  enough  to align the director field. 
Electric field  applied macroscopically 
is typically of order $1{\rm V/\mu m}$ \cite{P2,CL}.

\section{Simulation method}

We give our simulation method  in 
the Landau-de Gennes scheme 
under the  condition of $V=0$.  
For simplicity, we impose 
the periodic boundary condition 
in the $xy$ plane. 
We suppose  nanoscale particles confined 
between a thin layer.

\begin{figure}
\begin{center}
\includegraphics[scale=0.35,bb=0 0 418 336]{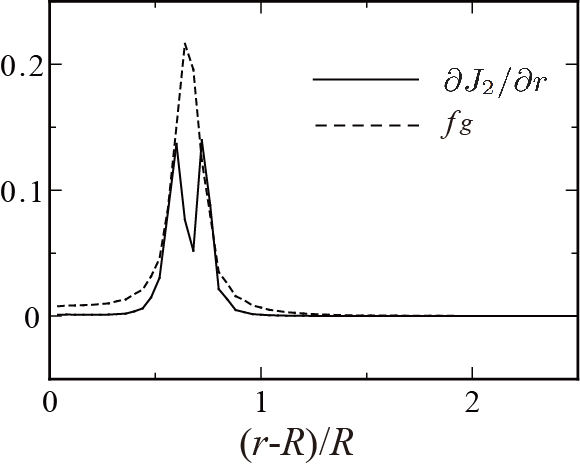}
\end{center}
\caption{
Derivative $\p J_2/\p r$ 
in  units of $d^{-1}$ 
and gradient free energy density $f_g= L(\nabla_k Q_{ij})^2/2$ 
in  units of  $Td^{-3}$   vs  normalized distance  
$(r-R)/R$ from the surface of  a charged 
spherical particle. The path starts from a surface position 
and passes through a Saturn ring (see Figure 3).   
}
\label{fig1}
\end{figure}

In the previous section we have assumed  
sharp boundaries  between the  particles 
 and the liquid crystal region. 
However, precise simulations are 
not easy  in the presence of 
 sharp curved boundaries  
on a  cubic lattice,  unless the mesh size is very 
small.    In this work, to overcome 
this difficulty, we employed  the smooth particle method. That is,   
we introduce  diffusive particle   profiles  
by \cite{Yama,Araki1,Araki2,Fuk04,Sta04}
\bea
\phi_n(\mbox{\boldmath $r$})=
\frac{1}{2}\tanh\left\{\frac{R-|\mbox{\boldmath $R$}_n
-\mbox{\boldmath $r$}|}{d}\right\}+ \frac{1}{2}, 
\label{eq:particle}
\ena
where the surface is 
treated to be diffuse with 
thickness    $d=T/L$ in equation (5),  
${\bi R}_n$ represents the particle center,  
and $R$ is the particle radius.

In terms of 
 $\phi_n(\mbox{\boldmath $r$})$, 
the overall particle and charge distributions are expressed as  
\bea
\phi(\mbox{\boldmath $r$})&=&\sum_n\phi_n(\mbox{\boldmath $r$}), \\
\rho(\mbox{\boldmath $r$})&=&
\frac{e}{v}\sum_n {Z_n}
\phi_n(\mbox{\boldmath $r$}), 
\ena
where $Z_n e$ are the particle  charges 
 and  $v=4\pi  R^3/3$ is the particle volume.  
The charge distribution is assumed to be homogeneous 
inside the particles. In $\mathcal{F}_0$ in 
equation (2) and $\mathcal{F}_g$ in equation (4), 
the  space integrals  outside the particles 
$\int' d{\bi r}$  should be redefined as 
\be 
\int ' d{\bi r}(\cdots)= 
\int d{\bi r}[1-\phi({\bi r})](\cdots).
\en 
The surface integral in equation (6) is also redefined as
\be
\int da(\cdots)=\int d\bi{r}|\nabla\phi|(\cdots).
\en
Then, the short-range anchoring 
free energy (6) is rewritten as 
\be
{\cal F}_a= -w 
\int d{\bi r}Q_{ij} (\nabla_i\phi)(\nabla_j\phi)
/|\nabla\phi|.
\en
The dielectric tensor is space-dependent as  
\be
\ve_{ij}(\mbox{\boldmath $r$})
=[\ve_0+(\ve_p-\ve_0)\phi]\delta_{ij}+
\ve_1(1-\phi )Q_{ij} , 
\label{eq:epsilon3}
\en
where $\ve_p$ is the dielectric constant 
inside the particles.

To seek $Q_{ij}$ satisfying equations (17) and (18), 
we treated $Q_{ij}({\bi r},t)$ 
as a time-dependent tensor variable 
obeying  the evolution equation, 
\bea
\frac{1}{\zeta} 
\frac{\partial }{\partial t} Q_{ij}({\bi r},t) 
&= &- \frac{\delta 
 \mathcal{F}}{\delta Q_{ij}}
+ \lambda\delta_{ij} \nonumber\\
&&\hspace{-2cm}= - (1-\phi)h_{ij} 
-L(\nabla_k\phi)(\nabla_kQ_{ij}) \nonumber\\
&&\hspace{-1.6cm}+ \frac{ w}{|\nabla\phi|} 
\bigg(
\nabla_i\phi\nabla_j\phi-|\nabla\phi|^2\frac{\delta_{ij}}{3}\bigg),
\ena 
where $\zeta$ is a constant kinetic coefficient. 
In the first line,  the functional derivative is taken 
both inside and outside the particles with the redefinitions  
(29)-(34), with   $\lambda$ ensuring  $Q_{ii}=0$. 
In the second line, 
$h_{ij}$ is defined in equation (16) and 
$ \nabla_k\phi$ arises from the factor 
$1-\phi$ in equation (31). 
On a cubic  $64\times 64\times 64$ lattice,  
 we integrated 
the above equation  for $Q_{ij}$. 
Space and time are measured in units of $d$ and 
\be 
\tau= d^2/\zeta L,
\en  
respectively. 
The space mesh size is $d$ and 
the  time mesh size is $\Delta t=0.01\tau$ in 
the integration. 
The cell interior is in the region 
$0 \leqq x,y,z \leqq 64d$. 
We   solved  the Poisson equation (14) 
at each integration step 
using a Crank-Nicolson method \cite{Foret06}.

As the boundary conditions  of $Q_{ij}$ at $z=0$ and $64d$, 
we assume  the homeotropic anchoring 
$n_i =\delta_{iz}$ for $\ve_1>0$ 
and the parallel alignment $n_i=\delta_{ix}$ 
for  $\ve_1<0$, where ${\bi n}=(n_x,n_y,n_z)$ 
is the director with $i=x,y,z$.  Those of $Q_{ij}$ 
in the $x$ and $y$ directions   are 
the periodic boundary conditions.  
The potential $\Phi$  vanishes 
at $z=0$ and $ 64d$ and is periodic in the 
 $xy$ plane. 
Note that the electric field 
at $z=0$ and $64d$ is along the $z$ axis, so 
the electrostatic energy is lowest 
for the selected director alignments both 
for $\ve_1>0$ and $\ve_1<0$. 
In order to approach a steady state, we 
 performed the integration  
 until $|d {\mathcal F}/dt| $ 
became less than $10^{-5}T/\tau$.

In our steady states thus attained, 
we confirmed  that  
both equations (17) and (18) excellently hold 
in the bulk liquid crystal region and near the 
particle surfaces, respectively. 
Mathematically, they should hold  in the thin-interface 
limit $d\ll R$,  where   $-\nabla \phi\cong 
\delta(r-R){\bi \nu}$ around a spherical surface 
 with $\bi \nu$ being 
the normal unit vector. 
In Figure 2, we  show our numerical result of 
the derivative 
$\p J_2/\p r= 2Q_{ij}\p Q_{ij}/\p r$ 
and the gradient free energy density 
$f_g = L(\nabla_k Q_{ij})^2/2$ around 
a particle surrounded by  a Saturn ring defect  for  $w=0$. 
See the next section for details of the calculation 
and Figure 3 for its 3D picture.  We can see that 
 $\p J_2/\p r$   is nearly equal to 
zero at the surface   
and exhibits double peaks 
around the  Saturn ring position. 
The boundary condition ${\bi \nu}\cdot\nabla Q_{ij}=0$ in 
equation (14) is thus nearly satisfied  
even in the presence of a  defect 
in our diffuse interface model.

\section{Numerical results}

In our simulations, we set     
\bea
&&A=-15T/d^3,\quad B=|A|/2, 
\quad C=3B,\nonumber \\
&& \ell_B= 12d, \quad 
\ve_p=\ve_0.\nonumber
\ena
For example, for $\ve_0= 2.3$ and $T=300$K, 
we have  $d=2$nm, $\ell_B=24$nm, and $L=2$pN. 
The nematic order parameter 
$S$ in  equation (20) 
is calculated as $S=0.75$. We show 
simulations results, where 
the charge number per  particle 
is $Z=30$, 50, 60, 80, 100, 
and 160.  If it  is  100 and 
the radius $R$ is 25nm, the surface electric field 
$E_s$ becomes $100$ V$/\mu$m.  
We  also set  $\ve_p=\ve_0$. 
In the case of one particle,   
the interior   dielectric 
constant $\ve_p$ does not affect the exterior 
electric potential and is irrelevant. 
In the case of  two particles, we 
also performed simulation with 
$\ve_p=2\ve_0$ in the examples in figures 6 and 8, 
but no marked difference was found.

In Subsections 4.1 and 4.2, 
we will neglect the short-range anchoring interaction 
and set  $w=0$, focusing  on the  electric field effect  
on the director field. 
In Subsection 4.3, we will include  the 
short-range anchoring interaction  around 
 a  charged particle.
In our  Landau-de Gennes scheme,  
the orientation order is    
almost uniaxial outside the defect cores   
both for  $\ve_1>0$ and  $\ve_1<0$.   
Thus we will display the 
director $\bi n$ around the particles. 
Tube-like  surfaces in Figures 4-10    will be those 
where $f_gd^3/T= (d\nabla_k Q_{ij})^2/2
=0.2$. This threshold is so high 
such that  the resultant tubes enclose   defects.

In addition, we confirmed that the eigenvalues of $\chi_{ij}$ 
in equation (22) were 
kept to be positive everywhere 
in the system. For example, in the uniaxial 
state with $\ve_1=1.8\ve_0$ and $S=0.75$,   
the eigenvalues of $\chi_{ij}$, 
are given by 
$\chi_{\parallel}\cong  0.27$ and 
$\chi_{\perp}\cong 0.024$. 

\subsection{A single particle in nematic liquid}

\begin{figure}
\begin{center}
\includegraphics[scale=0.5,bb=0 0 483 201]{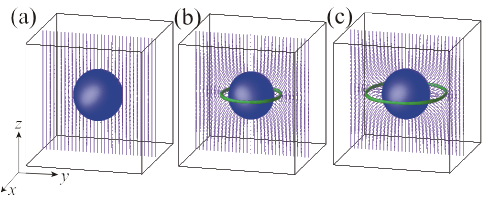}
\end{center}
\caption{
(color online) 
Orientational field  around a charged particle for  
(a) $Z=60$, (b) $Z=100$ and (c) $Z=160$ 
in a nematic solvent with $\ve_1=1.8\ve_0$. 
Short lines (in blue) 
 represent the director  ${\bi n}= 
(n_x,n_y,n_z)$  and 
cylinders   (in green)  in (b) and (c) 
contain a  Saturn ring.   
}
\end{figure}
\begin{figure}
\begin{center}
\includegraphics[scale=0.5,bb=0 0 492 196]{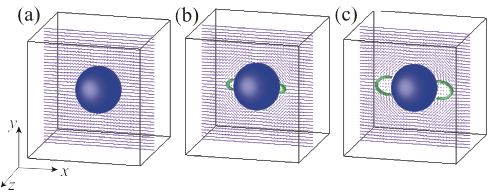}
\end{center}
\caption{
(color online) 
Orientational fields around a charged particle of 
(a) $Z=60$, (b) $Z=100$ and (c) $Z=160$ 
in a nematic solvent with $\ve_1=-1.8\ve_0$. 
}
\end{figure}

\begin{figure}
\begin{center}
\includegraphics[scale=0.8,bb=0 0 273 144]{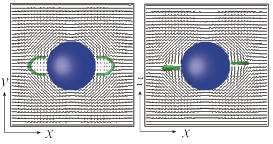}
\end{center}
\caption{
(color online) Top and side views 
of the director field ${\bi n}= 
(n_x,n_y,n_z)$  around ansae, corresponding to 
$Z=160$ in the panel (c) in Figure 4. 
}
\end{figure}

\begin{figure}
\begin{center}
\includegraphics[scale=0.4,bb=0 0 480 348]{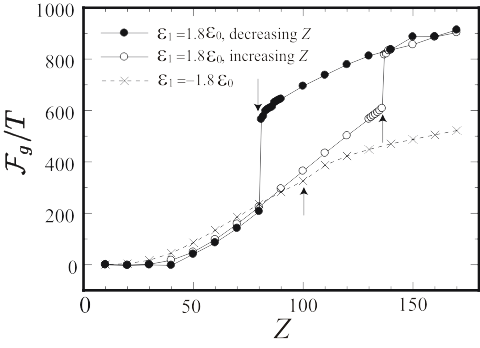}
\end{center}
\caption{Normalized gradient free energy $\mathcal{F}_g$  vs 
 charge number $Z$ of a particle. 
Arrows indicate  the point of defect formation. 
A jump appears  for $\ve_1>0$, 
while there is no jump 
for $\ve_1<0$.  
}
\end{figure}

We  fisrt consider a single charged particle 
for the two cases, $\ve_1>0$ and $\ve_1<0$. 
Its charge number $Z$ is 
in the range $[60,160]$. 
The orientation tensor $Q_{ij}$ is independent 
of the sign of $Z$.

Figure 3  displays the director field ${\bi n}=(n_x,n_y,n_z)$ 
around a  single particle with  $Z=60, 100$, and 160. 
Here we set $\ve_1=1.8\ve_0$, 
and  $R=12.5d$. 
The liquid crystal  is deeply in the nematic phase. 
At  the particle surface one of the  perpendicular 
alignment  is selected, 
which is analogous  
 to  the case of a neutral particle 
in the homeotropic anchoring condition realized for $w>0$. 
The system is axisymmetric, as assumed in our 
previous simulation\cite{Foret06}.  
For small $Z$ in (a),  
no defect is formed, while  the orientation 
 field  is largely distorted. 
For large $Z$ in (b) and (c), 
a Saturn-ring disclination line 
of the topological charge $s=-1/2$ 
appears near the equator of the particle. 
In our small system, the Saturn ring 
is confined within the box. However,  
if the system size is larger, 
the defect  should  be  more extended, 
since its radius is predicted to be of order $\ell$ in equation (24) 
\cite{Foret06}.

Figure 4 displays 
the orientation field  around a particle 
 with $R=12.5d$ for  $\ve_1=-1.8\ve_0$. 
The other parameters are the same as in Figure 3. 
For $\ve_1<0$, the director  tends to be along the 
particle surface, analogously to the case of 
a neutral particle with 
planar anchoring realized for $w<0$. 
For not large  $Z$  in (a),  
the director is distorted around the particle 
without defects. 
Slightly above the threshold in (b), 
defects are formed at the two poles of the particle. 
For a neutral particle, 
 a similar defect structure is 
 called ``boojum" \cite{Review,P1,Volovik}. 
For larger $Z$ in (c),  
 two ``ansa"-shaped defects emerge with their 
ends on  the particle surface, 
as a  novel defect structure. Here 
 a  boojum-like  structure in (b) grows 
into a curved disclination line of 
 topological strength $s=-1/2$. 
The director is perpendicular to  the 
plane formed by each ansa.  
In Figure 5 we show the top and side views 
of the director field around the ansae at $Z=160$ in Figure 4. 
Here the axial symmetry is broken, 
so the previous simulation did  not detect 
this structure \cite{Foret06} 
(where an axially  symmetric,  
biaxial  defect was instead detected).

For $\ve_1>0$ it was shown \cite{Foret06} that a Saturn-ring 
appears discontinuously with increasing 
$\ell (\propto Z)$ in equation (24). 
Also  in the case of 
a neutral particle \cite{Yama}, 
its appearance is discontinuous with 
increasing $wR$.  
In Figure 6,  we show  the normalized 
gradient free energy  
${\mathcal F}_g/T$ versus $Z$ for 
$\ve_1/\ve_0 =\pm 1.8$,  
since ${\mathcal F}_g$ in equation (4) 
is sensitive to the defect formation. 
The arrows indicate 
the point   of the defect formation on the curves. 
Remarkably, for $\ve_1=1.8\ve_0 >0$, ${\mathcal F}_g$  
jumps at $Z\cong 136$ with 
increasing $Z$ and at $Z\cong 80$ 
with  decreasing $Z$, where  $\ell/R 
\cong  5$ and 9, respectively,  using equation (24). 
This hysteretic behavior  demonstrates 
that  the system is bistable 
with and without  a Saturn ring 
(in the range of $80< Z < 136$ 
in the present example). 
On the other hand,  for $\ve_1<0$, 
$\mathcal{F}_g$ increases smoothly 
 as $Z$ increases.  This is because 
  the  ansa defects gradually protrude   
from the particle surface into 
the liquid crystal.

\begin{figure}
\begin{center}
\includegraphics[scale=0.68,bb=0 0 520 529]{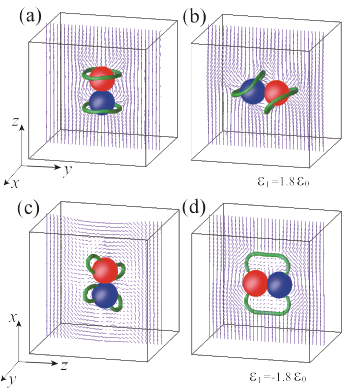}
\end{center}
\caption{
(color online) 
Director and   
defect structure around  oppositely charged particles 
 in the parallel direction 
(left) and in one of  the perpendicular directions (right) 
with respect to the background director direction. 
Upper plates (a)  and (b): 
 $\ve_1=1.8\ve_0$ and $Z_1=-Z_2=50$, 
where the free energy is lower for  (a) than for (b).  
Lower plates (c)  and (d): 
 $\ve_1=-1.8\ve_0$ and $Z_1=-Z_2=100$, where the free energy is 
lower for  (d) than for (c).  
}
\end{figure}

\subsection{A pair of charged particles}

\begin{figure}
\begin{center}
\includegraphics[width=0.41\textwidth,bb=0 0 520 529]{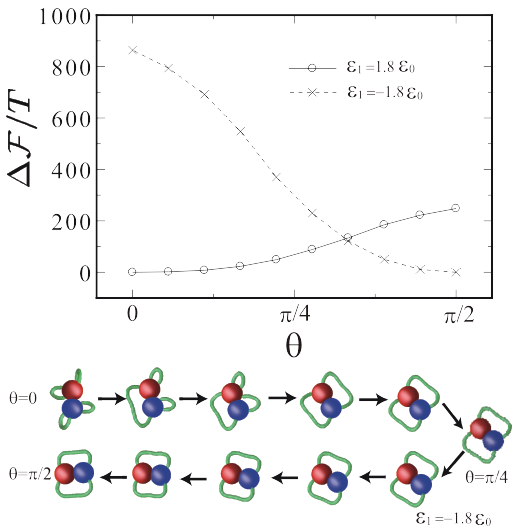}
\end{center}
\caption{(color online) 
Upper panel: 
Free energy difference 
$\Delta{\mathcal{F}}= {\mathcal{F}}(\theta)-
{\mathcal{F}}(\theta_{\rm min})$ 
for oppositely charged particles 
as  a function of  the angle $\theta$ between 
 the background director and 
the vector connecting the two  particles. 
Here $Z_1=-Z_2=50$ and 
$\theta_{\rm min}=0$ for the curve of  $\ve_1>0$, 
and $Z_1=-Z_2=100$  and 
$\theta_{\rm min}= \pi/2$ for the curve of $\ve_1<0$. 
Lower panel: 
Topological changes of the equilibrium  defect structure 
for $\ve_1=-1.8\ve_0$ for fixed 
$\theta= n \pi/20$  ($n=0, \cdots,10)$, 
corresponding  to the lower panels of  Figure 7. 
}
\label{fig7}
\end{figure}

\begin{figure}
\begin{center}
\includegraphics[scale=0.72,bb= 0 0 349 385]{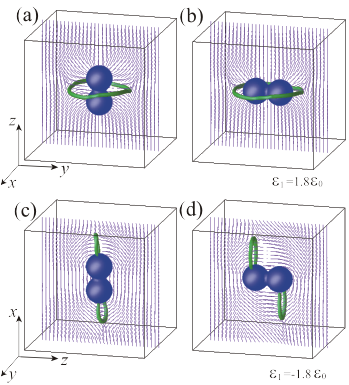}
\end{center}
\caption{(color online) 
Director and   
defect structure around  identically  charged particles  
 in the parallel direction 
(left) and in one of the perpendicular directions (right). 
Upper plates (a)  and (b): 
 $\ve_1=1.8\ve_0$ and $Z_1=Z_2=30$,  
where the free energy is lower for  (b) than for (a).  
Lower plates (c)  and (d): 
 $\ve_1=-1.8\ve_0$ and $Z_1=Z_2=80$, 
where the free energy is lower for  (c) than for (d). 
The defect topology  is the same as in the single 
particle cases in Figures 2 and 3.}
\end{figure}

\begin{figure}
\begin{center}
\includegraphics[width=0.38\textwidth,bb= 0 0 491 340]{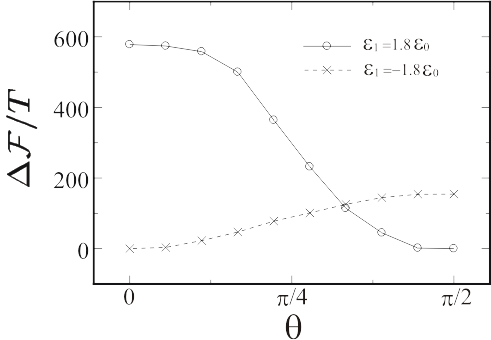}
\end{center}
\caption{
Free energy difference $\Delta{\mathcal{F}}= {\mathcal{F}}(\theta)-
{\mathcal{F}}(\theta_{\rm min})$ for 
identically   charged particles 
as a function of  $\theta$.  
 Here $Z_1=Z_2=30$ and 
$\theta_{\rm min}=\pi/2$ for the curve of  $\ve_1>0$, 
while $Z_1=Z_2=80$  and 
$\theta_{\rm min}= 0$ for the curve of $\ve_1<0$. 
 }
\end{figure}

We first  place  a pair of positively and negatively 
charged particles with $R=6.25$, which form a dipole. 
 Their  distance is fixed at 
 $|\mbox{\boldmath $R$}_1-\mbox{\boldmath $R$}_2|=2R$. 
In Figure 7, we show   snapshots 
of the director and the  defect structure 
around the two particles. Here 
$\ve_1=1.8\ve_0$ and $Z_1=-Z_2=50$ 
in the upper plates (a) and (b), while 
  $\ve_1= -1.8\ve_0$ and $Z_1=-Z_2=100$ 
in the lower  plates (c) and (d).
The particles are aligned in 
the parallel direction (left) 
and in one of the perpendicular  directions (right) 
with respect to the background director direction 
(along the $z$ axis  for $\ve_1>0$ and along the 
$x$ axis for $\ve_1<0$). 
We can see Saturn rings in (a) and (b), 
while there are four ansae in (c) and 
two ansae in (d). 
In the lower panel of 
Figure 8, we show the sequence 
of this topological change of the defect structure 
with varying the angle $\theta$ 
between the background 
director  and the vector connecting 
the particle centers. 
In the upper panel of 
Figure 8,  we show 
 the free energy ${\mathcal F}= {\mathcal F(\theta)}$ 
measured from its minimum $\mathcal F(\theta_{\rm min})$  
as a function of $\theta$. 
The   angle $\theta_{\rm min}$ at  the minimum is 
 $0$ for $\ve_1>0$  and $\pi/2$ for $\ve_1<0$.

We next place identically charged particles 
separated by $2R$.   
In Figure 9, we display the defect structures around 
two positively charged particle  with $Z_1=Z_2$. 
Remarkably, the topology  of the defects around a pair 
is the same as that  of a single particle. 
That is, we find  only one disclination loop 
for  $\ve_1>0$ and two  ansa  defects for $\ve_1<0$. 
Notice that a pair may be regarded as a non-spherical 
particle \cite{Hung} with charge $2Ze$. 
Figure 10 displays  the free  energy ${\mathcal{F}}
={\mathcal{F}}(\theta)$ 
measured from its minimum as a function of  
the angle $\theta$. 
The angle $\theta_{\rm min}$ at the  minimum 
is $\pi/2$ for $\ve_1>0$ 
and $0$ for $\ve_1<0$.

\subsection{A charged particle with nonvanishing  $w$}

In this subsection, 
we discuss the effect  of the short-range 
anchoring free energy  ${\mathcal{F}}_a$ 
in equation (6) supposing  a single 
particle. As illustrated  so far, the electric field for 
positive and negative $\ve_1$ serves to 
induce  homeotropic and planar alignment, respectively. 
Therefore,  the two anchoring mechanisms can compete 
for (i)  $\ve_1>0$ and $w<0$ and 
for (ii) $\ve_1<0$ and $w>0$.

\begin{figure}
\begin{center}
\includegraphics[width=0.45\textwidth,bb= 0 0 631 789]{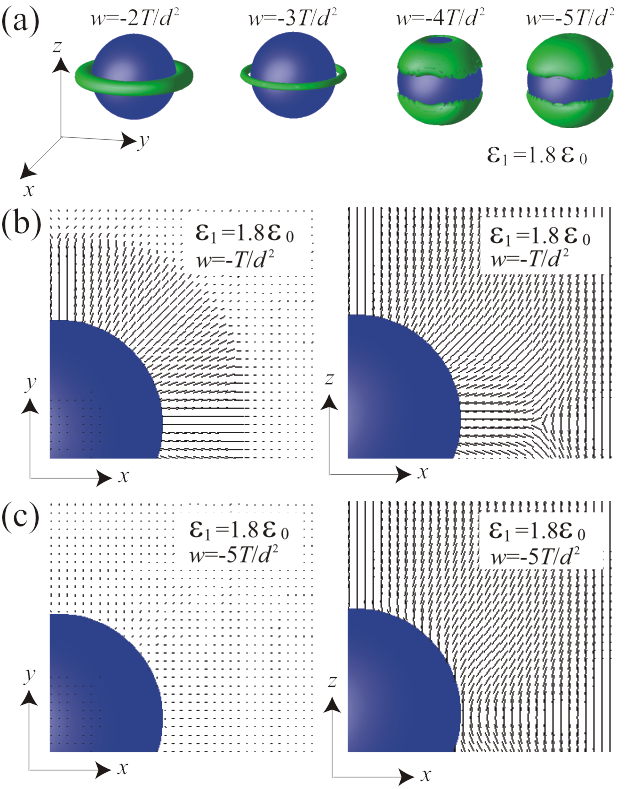}
\end{center}
\caption{ (color online) 
Results at  
$\ve_1=1.8\ve_0$  for negative $w$. 
Surface of   $f_gd^3/T=0.03$ 
for  $w d^2/T =-2,-3,-4$ and $-5$ in (a). 
Director $\bi n$ 
in the $xy$ plane ($z=32d$) (left) 
and in the $xz$ plane ($y=32d$) (right), 
where $w=-Td^{-2}$ in (b) 
and  $w=-5Td^{-2}$ in (c). 
The charge and radius 
of the particle are $Z=160$ and $R=12.5d$. 
}
\end{figure}

\begin{figure}
\begin{center}
\includegraphics[width=0.45\textwidth,bb=0 0 642 767]{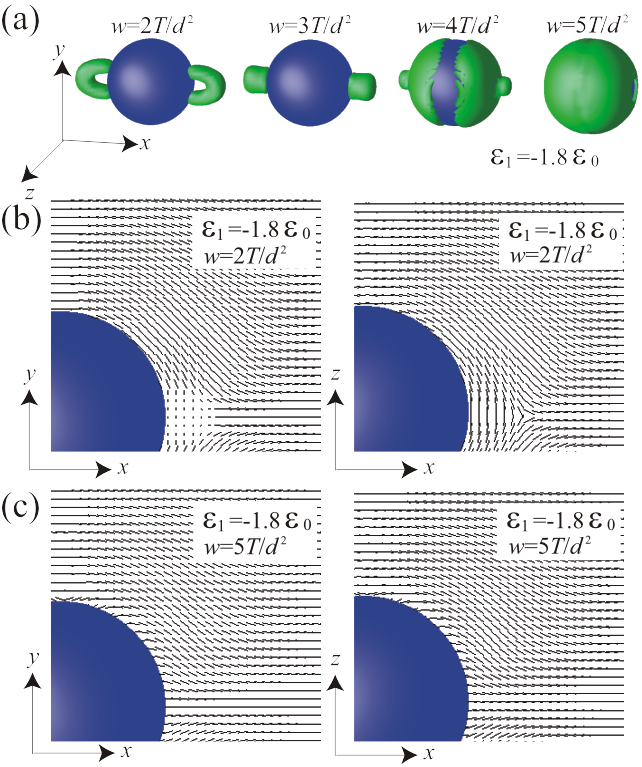}
\end{center}
\caption{(color online) 
Results at  $\ve_1=-1.8\ve_0$  for positive $w$. 
Surface of   $f_gd^3/T=0.03$ 
for  $w d^2/T =2,3,4$ and $5$ in (a). 
Director $\bi n$ 
in the $xy$ plane ($z=32d$) (left) 
and in the $xz$ plane ($y=32d$) (right), 
where $w=2Td^{-2}$ in (b) 
and  $w=5Td^{-2}$ in (c). 
Top and side views in (c) 
are indistinguishable.  
The charge and radius 
of the particle are $Z=160$ and $R=12.5d$. 
}
\end{figure}

In Figure 11, 
we set  $\ve_1=1.8\ve_0>0$  
and choose various negative $w$. 
In (a),   $f_gd^3/T =0.03$ on 
the surfaces (in green). This 
 threshold is small and the 
right two snapshots do not involve defects.  
We can see that the region having 
large $f_g$ moves from the vicinity of 
the Saturn ring  to  upper and lower surface parts  
of the particle.  The  Saturn ring  
remains   nonvanishing for small  $w$, 
but the director field around the equator  
tends to be tangential to the surface 
(parallel to the background director direction along  
the $z$ axis)  
and the Saturn ring  disappears 
with increasing  $|w|$. 
The director field 
changes steeply near the surface 
away  from the equator for large $|w|$. 
This changeover occurs 
discontinuously with sudden disappearance 
of the Saturn ring at 
 $w=w_c$, where  $w_c\cong -3.4Td^{-2}$ in the present case.

In Figure 12, 
we set  $\ve_1= -1.8\ve_0<0$  
and choose various positive  $w$. 
In (a), the ansa defects shrink into two point defects 
and disappear with  increasing  $w$. 
The top and side views of the 
director  are shown for $w=2Td^{-2}$ in  (b) and 
for $w=5Td^{-2}$ in  (c) around the particle.
For large $w$, 
the defect structure becomes  axisymmetric 
without defects 
and the regions of large $f_g$ 
 covers  the particle surface.
This crossover 
is  continuous with increasing $w$.

In the above examples,   
the short-range anchoring 
is effective close to the surface 
for sufficiently large  $|w|$, while 
the electric-field anchoring 
is dominant  far from the surface. 
A similar problem 
is encountered in  
the Fredericks transition in 
magnetic field as the 
strength of the surface anchoring 
is varied  \cite{Gen93,Rapini}.
The crossover from weak to strong 
short-range anchoring 
occurs for 
\be 
|w|> L/\xi_c=T/d\xi_c,
\en 
where $\xi_c$ is 
the thickness of this transition 
layer. For $|\ve_1|< \ve_0$ this length is  determined by 
\be 
 \xi_c^{-2}= E_s^2 |\ve_1|/8\pi LS ,
\en 
where $E_s=Ze/\ve_0R^2$ is the surface electric field. 
This estimation is obtained from $h_{ij}=0$ 
in equation (17). 
On the right hand sides of 
equation (16),  
 the gradient  term  becomes $-LS \nabla^2\varphi$ 
in the nematic phase,  
where $\varphi $ is the angle of the director with respect to
the surface normal. 
The balance of this term 
with  the last electrostatic 
 term ($\sim \ve_1 E_s^2/8\pi$) at   the surface 
yields equation (38). 
For our  parameters chosen in Figures 11 and 12, 
equation (38) gives  $\xi_c=0.91d$ 
and the right hand side of 
equation (37) becomes $1.1 Td^{-2}$, 
which are consistnt with our 
numerical results.

\section{Summary and Remarks}

We have performed three dimensional simulations 
in the presence of 
charged particles in   nematic liquid crystals. 
We first give a summary.\\ 
(i) 
The director 
tends to be parallel (perpendicular) to the electric field 
for positive (negative)  $\ve_1$. 
In Figure 3, a Saturn-ring defect  is  formed 
as $\ell$ in equation (24) much exceeds the particle radius $R$. 
In Figures 4 and 5, we have found  novel ansa defects 
without axial symmetry  in  
a nematic solvent with  $\ve_1<0$. 
In  our previous simulation \cite{Foret06}, 
a boojum-like  defect was derived for  $\ve_1<0$, 
since it was  based on the assumption of  axial 
symmetry.  In Figure 6, 
the formation of a Saturn ring 
due to electric field is first-order, while 
that of ansa defects is  continuous.\\
(ii)
We have also examined the director 
in the presence of two charged  particles  in nematic liquid crystals. 
Results for $Z_1=-Z_2$ are in Figures 7 and 8, while those 
for $Z_1=Z_2$ are in Figures 9 and 10. 
We have found that 
oppositely charged  particle pairs are 
 likely to be aligned 
in the parallel direction 
for $\ve_1>0$ and in  the 
perpendicular plane  for 
$\ve_1<0$ with respect to the background director direction. 
We conjecture that  polar molecules composed of oppositely 
charged particles  
can be aligned in nematic liquid crystals even on microscopic 
scales.  On the other hand, Figure 10 
shows that the preferred alignment 
directions are exchanged for identically charged particles. 
\\ 
(iii) We have examined competition of the charge-induced  
anchoring and the short-range anchoring 
in Figures 11 and 12. These two anchoring mechanisms can 
compete  when $\ve_1$ and $w$ 
have different signs. Under the condition (37), 
the short-range anchoring can be effective near the surface 
with distance shorter than $\xi_c$ in equation (38).\\

We supplement the discussion in Subsection 2.3. 
For microscopic particles (ions),  
 observation of  nanoscale 
defects should  be  difficult, but 
there might  be some indication 
of the defect formation in the behavior of 
the electric conductivity \cite{PG,Gen93}. 
For colloidal particles, 
 the condition $R> R_c$ 
can be satisfied only when the ionization on the surface 
occurs  to a sufficient level in  a liquid crystal. 
We  may also suspend a macroscopic 
particle in a liquid crystal. 
We mention an experiment \cite{P2}, in which 
an electric field was applied 
to   nematics  containing  
silicone oil particles to produce  
 field-dependent defects.  We may even propose 
to suspend metallic particles  or water droplets 
containing salt in a liquid crystal, where 
a surface charge appears in an applied  electric field.  
Recently,  electric field was applied to two-dimensional 
colloidal crystals in nematic solvent \cite{CL}, 
where the lattice spacing changes up to $20\%$  in one  direction 
in response to the applied field.

Further remarks are as follows.\\
(1) The competition  of the short-range and charge 
anchoring mechanisms should be investigated 
furthermore, since  our examples 
 of a single particle are very fragmentary.
The interaction among  charged  particles 
 in liquid crystal solvent should be much complicated 
in such situations.\\ 
(2) 
The liquid crystal order $S$ increases with 
increasing $|A|$ in the nematic phase and its discontinuity 
at the transition decreases with decreasing $B$ in equation (2). 
 For small $B$ (for weakly first order 
phase transition), therefore,  
the defect formation  takes  place considerably far  below the 
nematic-isotropic transition. 
The ion mobility in nematics   \cite{PG,Gen93} 
might  decrease discontinuously at the Saturn-ring formation 
with lowering the temperature.\\ 
(3) Light scattering  should be sensitive to doped ions 
 in nematics, where 
even a small amount of 
ions  should strongly distort the nematic order. 
This is  analogous  to  the role  of  microemulsions in nematics 
\cite{Yam01,Bel03}.\\
(4) 
As discussed below equation (26), 
we should examine how  ionic surfactant molecules 
can be attached to  surfaces of colloids and 
 microemulsions  in 
liquid crystal solvents \cite{OnukiEPL}. It is worth noting 
that an ionic surfactant was attached 
to microemulsion surfaces in the 
previous experiment \cite{Yam01}.\\ 
(5) 
Intriguing  also are  effects of salt  at weakly first order 
nematic-isotropic phase transition  and 
the ion distribution  around isotropic-nematic interfaces. 
Such theoretical studies 
 were already  reported for electrolytes 
with  binary mixture solvents 
 \cite{OnuKit04,OnukiPRE}. \\

\begin{acknowledgement}  
This work was supported by Grants-in-Aid 
for scientific research 
on Priority Area ``Soft Matter Physics" 
and  the Global COE program 
``The Next Generation of Physics, Spun from Universality and Emergence" 
of Kyoto University 
 from the Ministry of Education, 
Culture, Sports, Science and Technology of Japan. 
We thank Jun Yamamoto and Jan Lagerwall for valuable discussions. 
\end{acknowledgement}

\end{document}